\documentclass[%
 reprint,
 amsmath,amssymb,
 aps, 
 prl,
]{revtex4-2}

\usepackage{graphicx}
\usepackage{dcolumn}
\usepackage{bm}
\usepackage{color}
\usepackage[table,xcdraw]{xcolor}
\usepackage{multirow}
\usepackage{array}
\usepackage{booktabs}

\begin{document}

\preprint{APS/123-QED}

\title{Determining ground states of alloy  by a symmetry-based classification 
}


\author{Yu-Jie Cen}
\thanks{These authors contributed equally to this work}%
\author{Chang-chun He}
\thanks{These authors contributed equally to this work}%
\author{Shao-Bin Qiu}
\author{Yu-Jun Zhao}
\author{Xiao-Bao Yang}
\email{scxbyang@scut.edu.cn}
\affiliation{%
School of Physics and Optoelectronics, South China University of Technology, Guangzhou 510640, People’s Republic of China 
}%




\begin{abstract}
Reducing the number of candidate structures is crucial to improve the efficiency of global optimization. Herein, we demonstrate that the generalized Hamiltonian can be described by the atom classification model (ACM) based on symmetry, generating competent candidates for the first-principles calculations to determine ground states of alloy directly. The candidates can be obtained in advance through solving the convex hull step by step, because the correlation functions of ACM can be divided into various subspace according to the defined index $l$. As an important inference, this index can be converted to the number of Wyckoff positions, revealing the dominant effect of geometry symmetry on structural stability. Taking Ni-Pt, Ag-Pd, Os-Ru, Ir-Ru and Mo-Ru as examples, we not only identify the stable structures in previous theoretical and experimental results, but also predict a dozen of configurations with lower formation energies, such as Ag$_{0.5}$Pd$_{0.5}$ ($Fd$-$3m$),  Os$_{0.5}$Ru$_{0.5}$ ($Pnma$), Ir$_{1/3}$Ru$_{2/3}$ ($P6_{3}/mmc$), and Mo$_{0.25}$Ru$_{0.75}$ ($Cmcm$).


\end{abstract}

\maketitle


Determining the material structure theoretically  will provide important guidance to the experiment, where the energy evaluation and structure evolution are two main steps to predict the ground state structure.
 To explore configuration space, AIRSS \cite{ref3}, USPEX \cite{ref4}, CALYPSO \cite{ref5} have been successively developed based on the global optimization, which are widely used in the discovery of materials \cite{11,12,13,14}.
To improve the search efficiency, the energy evaluation can be simplified by the application of machine learning potential, which is trained from the data set based on density functional theory (DFT) \cite{28,29,30,39}.
For the alloy systems, the cluster expansion (CE) method uses the linear combination of basis functions to describe the Hamiltonian \cite{31}, and fits the interaction parameters through the data from DFT. 
To balance efficiency and accuracy, the effective cluster interactions should be selected for each specific material.
Based on genetic algorithm, a few of interaction parameters can be selected from a large number of interaction parameters, which can produce a good fitting of the Hamiltonian in some systems \cite{ref6}. In addition,  the parameters of CE can be  selected by using prior probability distribution combined with simple models or experience, based on the framework of Bayes' theorem \cite{ref7}.

Besides the simplification of energy evaluations,  reducing the number of candidate structures  is practical to accelerate global optimization.
Instead of scanning the interaction parameter space \cite{ref20},  the possible ground state structures of the system can be determined by finding convex hull in the feasible domain of the correlation function space by establishing geometric inequality constraints \cite{ref9,25,26,27}. 
By enumerating the structures of unique supercells of finite volume, the convex hull of the feasible region in the CE correlation function space can be directly solved to obtain the possible ground state structures of binary alloy \cite{ref10}. 
Based on simple geometric characteristics, the amount of the average bond type is used to evaluate the degree of deviation of the structure from a random situation, 
and  the candidates with a larger average bond type will be ground state structures \cite{ref8}. It is comprehensible that these structures are closer to the edge of the feasible region in the correlation function space, approaching the convex hull.

\begin{figure*}
	\centering 
	\includegraphics[width=0.95\textwidth]{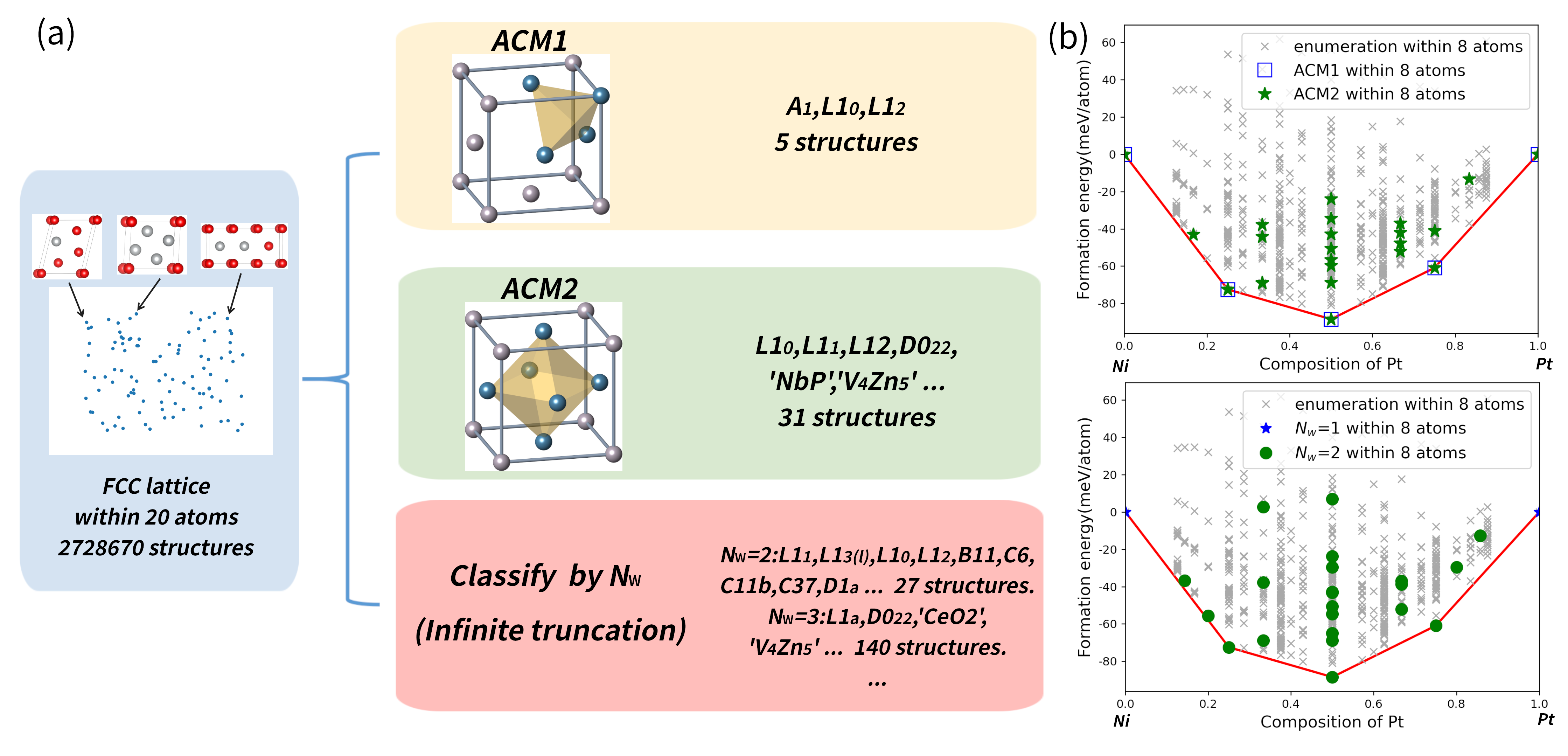}
	\caption{(a) The fcc-derivative structures within 20 atoms are filtered by ACM with various $R_{\rm cut}$ and the number of Wyckoff positions, and the order phases are obtained. (b) The formation energy of Ni-Pt alloy with cells within 8 atoms.}
\end{figure*}

Most previous studies focused on  the acceleration of energy evaluations with high accuracy combining the first-principles calculations and 
the interaction potential model \cite{28,29,30,39}. 
It is worth noting that the ground states of alloy are dominated by the energy model, in which the selection of parameters is closely related to the cut-off radii ($R_{\rm cut}$) and the training data set \cite{40}.
To screen candidates before energy evaluations, it will  become difficult to solve the convex hull in CE, when the dimensionality of the correlation function space increases with the increasing of interactions.
 For a set of $n$  points in $ \mathbb{R}^{d} $, the regular triangulations algorithm takes $ O(n\, log \, n + n^{d/2}) $ time \cite{38,ref22}, making it  impossible to achieve medium-sized inputs when the function space exceeds 10 dimensions.

In this letter, we show that the ground state structures of alloys can be directly obtained by the first-principles calculations, where the candidates are screened by the symmetry-based classification. Firstly, we demonstrate that atom classification model (ACM) in our previous study \cite{37} is equivalent to CE considering the complete many-body interaction, and the classified atoms correspond to Wyckoff positions for the infinite $R_{\rm cut}$. Secondly, we propose an effective method for calculating the convex hull with a given $R_{\rm cut}$, and the convex hull can be solved step by step in a certain order by dividing the correlation functions of ACM into various subspace.
Finally, 
we provide a high-throughput first-principles calculations strategy to explore the stable structures of alloys, reducing the number of candidates significantly.



As shown in Fig. 1(a), there are 2728670 unique configurations of $A_{x}B_{1-x}$  within 20 atoms in the fcc lattice. With the truncation of tetrahedron, there are five possible ground state structures according to the convex hull of ACM. These structures are the typical ordered phase ($L1_{0}$ and $L1_{2}$) found in various binary alloys, such as  Cu$_{x}$Au$_{1-x}$ and Ni$_{x}$Pt$_{1-x}$. The number of vertex structures will increase to 31 and more ground state structures in experiment are conformed, when the truncation is expanded to the  octahedron. Meanwhile, most ordered alloy structures in the fcc lattice can be obtained by considering the candidates with the number of Wyckoff positions ($N_{\rm W}$) no more than three. 
In the case of  Ni$_{x}$Pt$_{1-x}$ [shown in Fig. 1(b)], all the five ground state structures can be determined after the calculations of candidates obtained by ACM with the truncation of tetrahedron. The first-principles calculations details are shown in supplemental material (SM.) Sec.1. For a larger truncation of octahedron, no new stable structures are found. Similarly,  the five ground state structures can be reproduced in 29 candidates with $N_{\rm W}\leq 2$ and no new stable structures are uncovered in the 140 candidates with $N_{\rm W}=3$.


Generally, the ground state structures of the material are highly symmetric in geometry. In the following, we show $N_{\rm W}$ is the key index to describe the structural stability and the symmetry, through a concise demonstration based on ACM. Taking the one-dimensional atomic chain  as an example [as shown in Fig. 2(a)], we construct the linear Hamiltonian based on the original lattice-gas model (LGM) \cite{32}, CE, and ACM as follows:  
\begin{equation}
    H = \bm{J}\cdot \bm{\overline{\Pi}}
\end{equation}
where $\bm{J}$ is the interaction parameter, and $\bm{\overline{\Pi}}$ is the correlation function shown in Table I. The deduction of ACM is in SM. Sec. 2.1.


\begin{table}[h]
\renewcommand\arraystretch{1.7}
 \resizebox{\linewidth}{!}{
\begin{tabular}{ccc}
\multicolumn{1}{l}{\cellcolor[HTML]{FFFFFF}{\color[HTML]{333333} }} &
  J &
  $\overline{\Pi}$ \\ \hline
 &
  $J_{1}=\mu_{A},$ &
  $\Pi_{1}=\frac{1}{N}\sum_{i} P_{i}^{A},$ \\
 &
  $J_{2}=\mu_{B},$ &
  $\Pi_{2}=\frac{1}{N}\sum_{i}P_{i}^{B},$ \\
 &
  $J_{3}=J_{AA},$ &
  $\Pi_{3}=\frac{1}{N}\sum_{i,j}P_{i}^{A}P_{j}^{A},$ \\
 &
  $J_{4}=J_{AB},$ &
  $\Pi_{4}=\frac{1}{N}\sum_{i,j}(P_{i}^{A}P_{j}^{B}+P_{i}^{B}P_{j}^{A}),$ \\
\multirow{-5}{*}{LGM} &
  $J_{5}=J_{BB}$ &
  $\Pi_{5}=\frac{1}{N}\sum_{i,j}P_{i}^{B}P_{j}^{B}$ \\ \hline
 &
  $J_{1}=\frac{1}{2}(J_{AA}-J_{BB}+\mu_{A}-\mu_{B}),$ &
  $\Pi_{1}=\frac{1}{N}\sum_{i}S_{i},$ \\
 &
  $J_{2}=\frac{1}{4}(J_{AA}-2J_{AB}+J_{BB}),$ &
  $\Pi_{2}=\frac{1}{N}\sum_{i,j}S_{i}S_{j},$ \\
\multirow{-3}{*}{CE}  & $J_{3}=\frac{1}{4}(J_{AA}+2J_{AB}+J_{BB}+2\mu_{A}+2\mu_{B})$ & $\Pi_{3}=1$                                           \\ \hline
 &
  $J_{1}=\mu_{A}+J_{AA},$ &
  $\Pi_{1}=\frac{1}{4N}\sum_{i,j}(1+2S_{i}+S_{i}S_{j}),$ \\
                      & $J_{2}=\frac{1}{2}(\mu_{A}+\mu_{B}+J_{AB}),$                 & $\Pi_{2}=\frac{1}{4N}\sum_{i,j}(2-2S_{i}S_{j}),$      \\
\multirow{-3}{*}{ACM} & $J_{3}=\mu_{B}+J_{BB}$                                       & $\Pi_{3}=\frac{1}{4N}\sum_{i,j}(1-2S_{i}+S_{i}S_{j})$
\end{tabular}}
\caption{The components of $J$ and $\overline{\Pi}$ in LGM,CE and ACM.}
\label{tab:my-table}
\end{table}


In the representation of CE and ACM [see Figs. 2(b) and 2(c)], we have determined the possible ground state structures of the system by finding the vertices in the feasible region of the correlation function space, by enumerating all the structures of the one-dimensional atomic chain of binary alloys within 8 atoms. With the transformation matrix [see Fig. 2(b)], the correlation functions of ACM and CE can be converted to each other, confirming the equivalence of these two methods in energy evaluation in the framework of LGM. In addition, we have demonstrated that ACM is equivalent to CE when  the complete many-body interactions are considered. For the two-dimension square lattice truncating to next-neighbor interaction, there are 20 kinds of interactions in LGM  and the system can be equivalently represented by 6 kinds of four-body interactions in ACM, where all the 2,3-body interactions are included according to the deduction (see SM. Sec. 2.2).

Although more parameters are required in ACM compared with CE, 
the correspondence between the cluster environment in ACM and the structure is clearer, revealing the relation of structural stability and the symmetry of configuration. More importantly, with the benefit of the non-negative components of correlation functions, ACM brings convenience for the subsequent calculation of convex hulls to find possible ground state structures.

Mathematically, the vertices of a set is defined as follows: in a point set $X=$ \{$\bm x_1$, $\bm x_2$, $\bm x_3$, ...\} with any point $\bm x_{i}$ belonging to $X$,  the convex set is marked as $Y=$ \{$\bm y_{1}$, $\bm y_{2}$, $\bm y_{3}$, ..., $\bm y_{m}$\} ($m$ vertices in total), satisfying

	\begin{equation}
		k_{1}\bm{y}_{1}+k_{2}\bm{y}_{2}+...+k_{m}\bm{y}_{m}=\bm{x}_{i}
	\end{equation}
with $ \sum_ {i} k_ {i} = 1 $, $ k_ {i} \geq 0 $.
Note that the ACM correlation function $\overline{\Pi}=(\Pi_{1},\Pi_{2},...,\Pi_{n})$ satisfies $\sum_{i}\Pi_{i} = 1$ and $\Pi_{i}  \geq 0$, indicating that all data points will be constrained in a finite $n-1$ dimensional hyperplane.

To ensure that all possible ground state structures are included, 
we can enumerate supercells of various sizes and shapes until the convex hull no longer changes. 
As shown in the left of Fig. 2(b), there are only 3 structures by screening the candidates within 8 atoms with the convex hull in CE, and the corresponding coordinates are (0, $-1$), ($- 1$, $-1$), (1, 1). However, we cannot be sure that whether the convex hull will change in larger cells. 
By contrast, we find 3 convex hull structures in ACM, i.e., (1, 0, 0), (0, 1, 0), (0, 0, 1) respectively as shown in the right of Fig. 2(b). They are located on the axis of correlation function space.  It can be concluded that the current convex hull are complete, and  the convex hull will remain the same even searching larger cells.
Note that these structures on the axis are composed of only one type of cluster, exhibiting high symmetry in geometry.
\begin{figure}[h]
	\centering
	\includegraphics[width=0.48\textwidth]{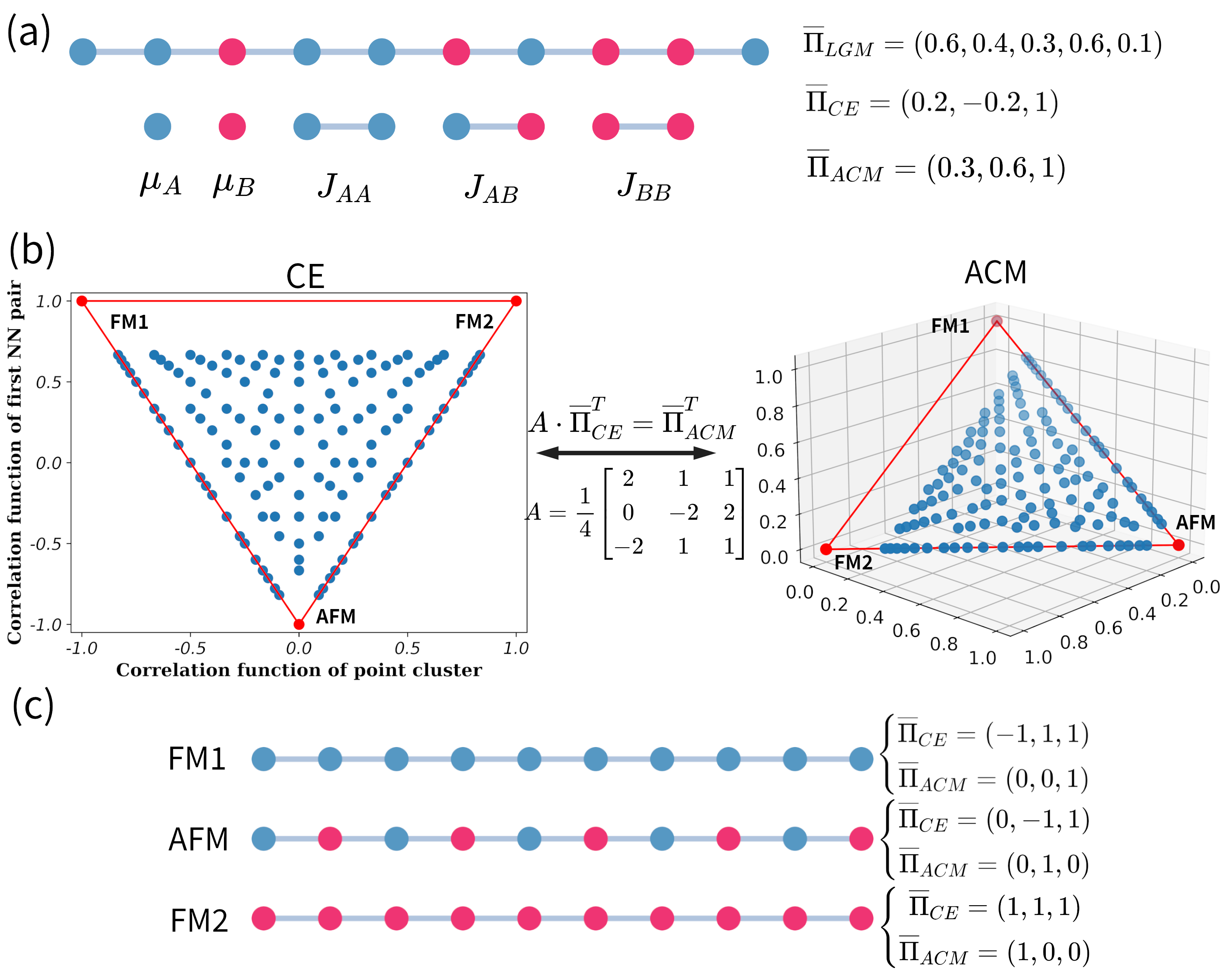}
	\caption{(a) On the left side is a random periodic one-dimension atomic chain and 5 interaction type. On the right side is the correlation functions in LGM, CE and ACM. (b) The correlation functions space of CE and ACM. (c) Three convex hull structures and their corresponding correlation functions.}
\end{figure}

In the framework of ACM, there is a key index $l$, i.e., the number of non-zero components of correlation functions. The convex hull is complete if the number of  structures with $l=1$ is equal to the dimensionality of space, because all the possible candidates can be linearly represented by these structures. Besides the 1D atom chain (shown in Fig. 2), there are three additional cases: 2D square lattice, 2D triangular lattice, and 3D fcc lattice with the certain $R_{\rm cut}$ (see SM. Sec. 4.1-4.3).

When the number of structures with $l=1$ is less than the dimensionality of space, we can classify the structures according to the dimensionality of $l$, because all the data points in the ACM correlation function space are restricted to a finite hyperplane, and those with larger  $l$ are distributed inside the hyperplane.
Note that $\overline{\Pi}_{\rm ACM}$ with smaller $l$  can not be linearly represented by the ones with larger $l$ and the convex hull can be solved in the order of $l$ increasing sequences, making the search becomes memorable.
Furthermore, we can decompose the high-dimensional correlation function space into a combination of subspaces according to  non-zero components of the correlation functions, and determine the vertices in each subspace gradually.

Based on the above ideas, we have designed an algorithm from generating structures to calculating convex hull, with the flow chart shown in Fig. 3.
We propose a ``jigsaw puzzle" to directly generate structures with small $l$ to improve the search efficiency.
Beginning from one position in a given cell, the clusters gradually are added outwards, 
 and the selection of cluster environment is arbitrary permutation and combination under the restriction of the ACM generating group.
Finally, the puzzle is completed and the duplicate structures are filtered out based on symmetry, clearly minimizing the cost of producing candidates. The details about the algorithm are described in SM. Sec. 3.

\begin{figure}[h]
	\centering 
	\includegraphics[width=0.48\textwidth]{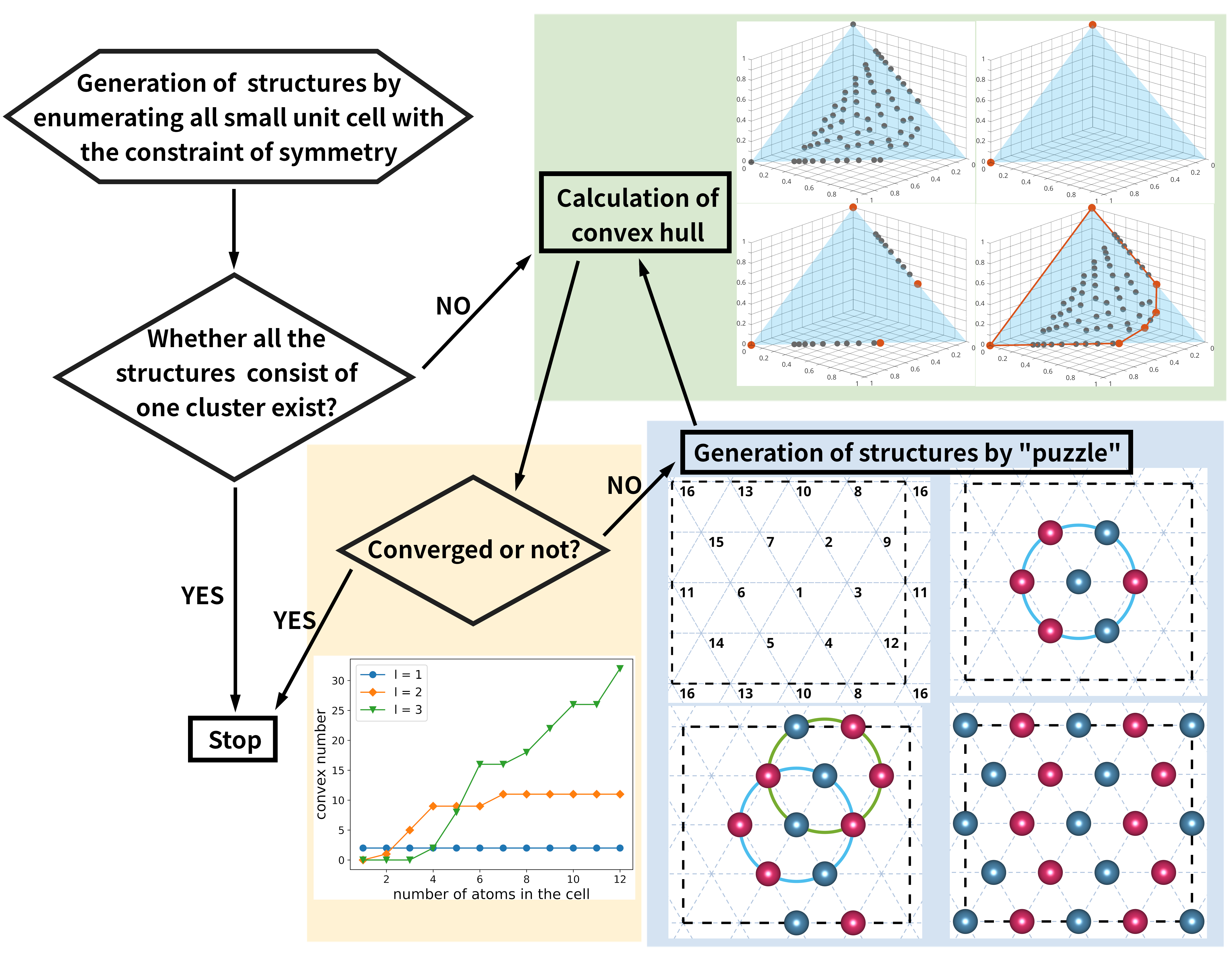} 
	\caption{The flow chat of determining the convex based on ACM.}
\end{figure}

 Taking  the two-dimensional triangular lattice as an example, we study the evolution of ordered structures in the binary alloy system ($A_xB_{1-x}$) as the $R_{\rm cut}$ increases (see SM. Sec. 4.2). When the truncated environment of 3 atoms [see Fig. S14(a)] is selected, there are 4 types of unique atomic clusters and  4 vertex structures corresponding to $l=1$, indicating a complete convex hull in this case. Selecting a truncated environment with 7 atoms, the number of unique atomic clusters  increases to 26, and only the pure phases containing only $A$ or $B$ atoms correspond to $l=1$. 
 According to the above algorithm, we have determined all possible vertex structures with $l=2$. 
 As shown in Fig. S14(b), the structures $b_{6}$, $b_{7}$, and $b_{3}$ correspond to the ordered structures ($ \sqrt{3} \times \sqrt{3} $), ($2 \times 2 $), and ($ \sqrt{7} \times \sqrt{7} $) that alkali metal atoms adsorbed on graphite in the experiments \cite{33}. As shown in Figs. S14(c) and S14(d), the structure $b_{3}$ is not a vertex structure when the truncated environment is  3 atoms, becasue it can be linealy expressed by the structures of $a_{2}$ and $a_{4}$ with $l=1$. With the truncation environment of 7 atoms, $a_{4}$ is convex with $l=1$, while the $l$ of vertex structure $a_{2}$ increases from 1 to 2. As a result, the structure $b_{3}$ becomes an additional vertex structures with $l=2$ in the new correlation function  space. 



It can be comprehensible that  the degenerate structures in the correlation function space with smaller $R_{\rm cut}$ will be distinguished by the increasing of $R_{\rm cut}$, and new ground states will be uncovered. The number of non-zeros component $l$ of the structure may increase with the $R_{\rm cut}$, but the upper limit of $l$ is the number of Wyckoff sites of the structure, corresponding to the infinite $R_{\rm cut}$.
According to the convex hull of ACM, the candidates with small $l$ will be ground states with more probability, where smaller $l$ corresp higher symmetry of the crystal structure. 
This explains very well why stable structures are usually considered to be highly symmetrical, simply geometrical, containing few atoms \cite{ref8}.
Counter-intuitively, structures with small $N_{\rm W}$  do not only exist in unit cells with few atoms. For example, there is a unit cell of 16 atoms in a fcc lattice with only $N_{\rm W}=2$ (structures that cannot be reduced to smaller unit cell). 
However, it is still difficult to generate all the possible candidates with small $N_{\rm W}$ with the increasing of cell size.

In our previous studies, the unique configurations are obtained by structural recognition \cite{34}.
To give a further screening of candidates, we provide two ways to calculate the ground state structures of the alloy: i) choose the ACM model with various $R_{\rm cut}$ and generate the vertex structures for the first-principles calculations; ii) classify the structures by $N_{\rm W}$, and gradually perform the calculations follow the order of $N_{\rm W}$ increasing sequences. Fig. 4(a) shows the efficiency of the two strategies, where the fcc-derivative structures are reduced from $10^6$ to $10^2$-$10^3$ for the unit cells within 20 atoms.

\begin{figure}
	\centering 
	\includegraphics[width=0.48\textwidth]{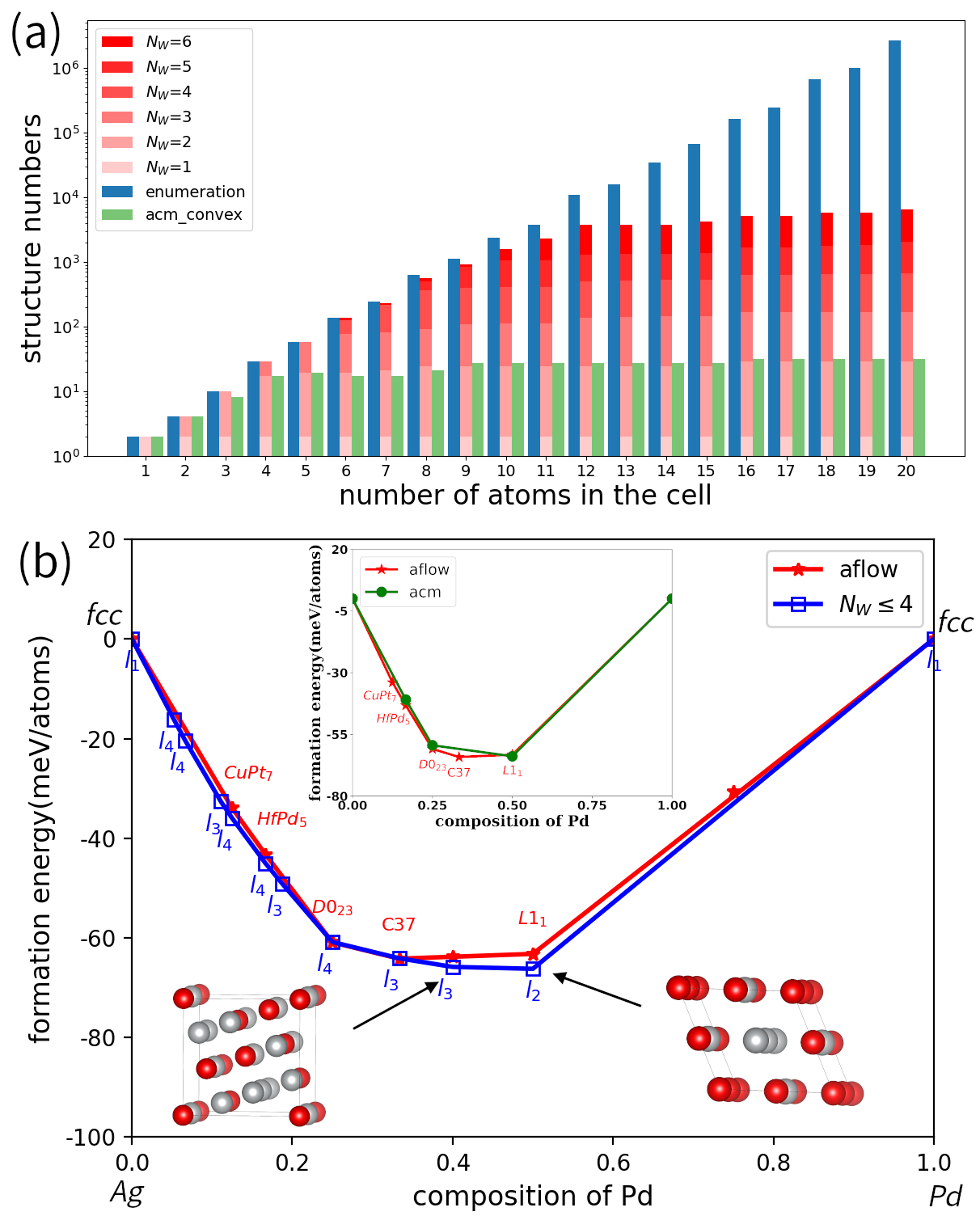} 
	\caption{(a) The comparison of the number of fcc-derivative structures between enumeration, 6-atom truncation ACM and $N_{\rm W}$ classifications method. (b) The convex hulls of formation energies for the Ag-Pd system. $l_2$, $l_3$, $l_4$ indicate the structures with $N_{\rm W}=2, 3, 4$, respectively.} 
\end{figure}



In the following, we will combine these two strategies with the first-principles calculations on alloys of Ag-Pd, Os-Ru, Ir-Ru and Mo-Ru, and compare with previous high-throughput calculations \cite{41}. The candidates generation details are shown in SM. Sec. 7.
Fig. 4(b) shows the formation energies of Ag-Pd (the results of Os-Ru, Ir-Ru and Mo-Ru systems are shown in Fig. S17) 
with the candidates from our two strategies, and the comparison with the previous calculations is demonstrated.
In the first strategy, that is, screening structures by ACM with a moderate $R_{\rm cut}$, we only need to calculate a small amount of the structures and we can revisit the main results of AFLOW. For example,  we only calculated 31 fcc-derivative structures for Ag-Pd [see inset of Fig. 4(b)], 125 hcp-derivative structures for Os-Ru,  156 structures (31 fcc- and 125 hcp-derivative) for Ir-Ru (see Fig. S17 in SM. Sec. 7.).

According to the second strategy, the structures with $N_{\rm W} \leq 4$ are calculated. Despite of the higher computational cost with more structures, we reveal several as-yet unreported ordered states of Ag-Pd, Os-Ru,Ir-Ru and Mo-Ru, whose formation energies are lower than those from AFLOW's results.
Based on the calculations of 673 superstructures with fcc-derivative, there is a  Ag$_{0.5}$Pd$_{0.5}$ structure with $Fd$-$3m$ symmetry which is more stable than $L1_{1}$, containing a unit cell of 8 atoms with $N_{\rm W}=2$. Another stable Ag$_{0.6}$Pd$_{0.4}$ structure ($P4_{2}/n$) is found, with a unit cell of 20 atoms and $N_{\rm W}=3$. [see Fig. 4(b)]
Among the calculated 436 hcp-derivative structures of Os$_{1-x}$Ru$_{x}$, a new structure of Os$_{0.5}$Ru$_{0.5}$ ($Pnma$) was found to be more stable than $B_{19}$, and there are 16 atoms in its unit cell with $N_{\rm W}=4$. The structure of Os$_{1/3}$Ru$_{2/3}$($Cmcm$) is also stable according to the convex hull of formation energy, corresponding to a unit cell of 6 atoms and $N_{\rm W}=2$. [see Fig. S17(a)]

For Ir$_{1-x}$Ru$_{x}$, we have performed the calculations of 673 fcc-derivative structures and 436  hcp-derivative structures. A new phase of Ir$_{1/3}$Ru$_{2/3}$ ($P6_{3}/mmc$) is found to be more stable than the phase of $Ir_{2}Tc$, possessing a unit cell of 18 atoms with $N_{\rm W}=3$ [see Fig. S17(b)].
Based on the calculation of 682 bcc-derivative structures and 436  hcp-derivative structures, the most stable Mo$_{0.25}$Ru$_{0.75}$ corresponds to the bcc lattice, whose energy is very close to the hcp-derivative from AFLOW's results. This new Mo$_{0.25}$Ru$_{0.75}$ structure possesses the symmetry of $Cmcm$, with 8 atoms in the unit cell and $N_{\rm W}=3$. In addition, Mo$_{0.125}$Ru$_{0.875}$ ($P$-$6m2$) is found to be stable,  with a unit cell of 8 atoms and and $N_{\rm W}=4$. [see Fig. S17(c)]

In summary, we demonstrate that ground states of alloy can be feasibly determined based on ACM,  revealing the connection between stable alloys and geometric symmetry, i.e., the structures with lower $l$ or $N_{\rm W}$ will become ground states with greater probability. 
The number of candidates for the first-principles calculations is remarkably reduced. According to the formation energies of Ag-Pd, Os-Ru, Ir-Ru and Mo-Ru systems, we have predicted many configurations with higher stability. Our finding provides a standard routine to screen stable alloy structures, facilitating high-throughput first-principles calculations of alloys.

\begin{acknowledgments}
This work was supported by Guangdong Basic and Applied Basic Research Foundation (Grant No. 
2021A1515010328), Key-Area Research and Development Program of Guangdong Province (Grant No. 
2020B010183001), and National Natural Science Foundation of China (Grant No.12074126).
\end{acknowledgments}

\bibliography{apssamp}

\end{document}